\documentstyle{aipproc}
\begin{document}
\title{An Efficient Matched Filtering Algorithm for the Detection of Continuous Gravitational Wave Signals}
\author{Peter R. Williams and Bernard F. Schutz}
\address{MPI for Gravitational Physics, Albert Einstein Institute, Am M\"{u}hlenberg 1,\\D--14476 Golm, Germany}
\maketitle
\section*{Introduction}
Neutron stars are perhaps the most promising class of gravitational wave (GW) sources, and searches for such GW signals is particularly 
suited to the characteristics of the GEO600 detector (see Schutz ``Getting Ready for GEO600 Data'' gr--qc9910033). However, the 
instantaneous GW frequency of such a source will evolve due to both intrinsic spindown effects and Doppler modulations induced by the 
motion of the Earth. Thus because of the large parameter space of likely signals, directly implemented optimal matched filtering is not 
computationally feasible.

In response to this problem, Schutz and Papa have developed an alternative strategy: the Hough--Hierarchical search 
algorithm (see Schutz and Papa ``End--to--End Algorithm for Hierarchical Area Searches for Long--Duration GW Sources for GEO600'' gr--qc9905018). 
In order to carry out a blind search over a range of intrinsic GW frequencies, the following three stages must be 
calculated for each point in the parameter space of sky positions and intrinsic spindown parameters:
\begin{description}
\item Stage I: Calculate demodulated Fourier transforms (DeFTs) on an intermediate time baseline (of order 1 day) by combining FFTs of 
short durations (approximately 30 minutes) of the time series data. In this context demodulated means that if there is a source at 
the sky position in question, and with the intrinsic spindown parameters in question, then all spindown and modulatory effects will have been correctly 
removed from the DeFTs: all signal power will be confined to one and the same frequency bin in each DeFT. This frequency is the intrinsic 
frequency of the source measured at the start of the observing time. It is expected that the total observing time will be of order 4 months, and 
thus roughly 120 of these DeFTs will be calculated for each point in parameter space.

\item Stage II: In general source parameters will not coincide exactly with those searched for, and residual frequency evolution and modulation will 
remain in the DeFTs. Thus, the peak in power associated with a given source may change frequency bins from DeFT to DeFT. Because of the 
relatively small time baseline of these DeFTs and the resultant poor signal--to--noise of any expected continuous GW signal, this evolution 
will be not directly apparent in the DeFTs, but can be recovered statistically using the Hough Transform algorithm.

\item Stage III: Calculate DeFTs for candidate sources with the full frequency resolution of the total observation time, by combining the intermediate 
baseline DeFTs produced in stage I.
\end{description}
Thus, during stage II, regions of the parameter space in which it is statistically unlikely that there are GW sources are eliminated from the search. 
Thereby, in stage III, the most computationally expensive part of the algorithm, the long time baseline DeFTs are calculated over only a very small 
fraction of parameter space and over a very small range of frequencies.

In this paper we outline the methods used in the first and third stages of this algorithm in constructing a longer time baseline DeFT from a number of 
shorter time baseline FFTs or DeFTs.

\section*{The Method}

Consider a time series $x_{a}$ of total duration $T$, which has been divided into $M$ short time series, each having 
$N$ data points. Then the DeFT for a signal with a time independent amplitude and phase $2\pi\Phi_{ab}(\vec{\lambda})$ is
\begin{equation}
\hat{x}_{b}(\vec{\lambda}) =  \sum_{a=0}^{NM-1}x_{a}e^{-2\pi i \Phi_{ab}(\vec{\lambda})} = \sum_{\alpha=0}^{M-1}\sum_{j=0}^{N-1}
x_{\alpha j}e^{-2\pi i \Phi_{\alpha j b}(\vec{\lambda})}, \label{mat}
\end{equation}
where the time indices are related by $N\alpha+j=a$, and $b$ is a long time baseline frequency index. In the following discussion Latin indices $j,k,l$ always 
sum over $N$, while Greek indices sum over $M$. Note that $\Phi_{ab}(\vec{\lambda})$ is dependent 
on a vector $\vec{\lambda}$ of parameters which characterize the signal one is searching for. In searching for GW signals from neutron stars these will 
include intrinsic spindown parameters, and the position of the source in the sky. If $\tilde{x}_{\alpha k}$ is the matrix formed by carrying out 
Fourier transforms along the short time index $j$ in $x_{\alpha j}$, then equation \ref{mat} can be written as
\begin{equation}
\hat{x}_{b}(\vec{\lambda}) = \sum_{\alpha=0}^{M-1}\sum_{k=0}^{N-1}\tilde{x}_{\alpha k}\,\left[\frac{1}{N}\sum_{j=0}^{N-1}
e^{-2\pi i\left( \Phi_{\alpha j b}(\vec{\lambda})-\frac{jk}{N}\right)}\right]
= \sum_{\alpha=0}^{M-1}Q_{\alpha}(b,\vec{\lambda})\sum_{k=0}^{N-1}\tilde{x}_{\alpha k}P_{\alpha k}(b,\vec{\lambda}), \label{deft}
\end{equation} 
where the product $Q_{\alpha}(b,\vec{\lambda})P_{\alpha k}(b,\vec{\lambda})$ is defined by the terms in square brackets, and 
$Q_{\alpha}(b,\vec{\lambda})$ contains all parts of the square brackets independent of the short time index $j$ and short frequency index $k$.

In equation \ref{deft} we have effectively re--written equation \ref{mat}, a long time baseline DeFT in the time domain, as a sum 
($\alpha$ index) of short time baseline DeFTs in the frequency domain ($k$ index), where $Q_{\alpha}(b,\vec{\lambda})P_{\alpha k}(b,\vec{\lambda})$ are these 
frequency domain filters. In the presence of stationary noise with a flat spectrum, equation \ref{deft} is the optimal 
detector. However, through applying various approximations, the detector can be made ``acceptably sub--optimal'', in the sense that only a small 
fraction of power from a signal is lost in comparison to the optimal case, while achieving vast savings in computational cost.

To illustrate these mathematical approximations it is instructive to discuss equation \ref{deft} for a specific case of 
 $\Phi_{\alpha j b}(\vec{\lambda})$: a linearly varying frequency model, i.e. in the continuum limit $\Phi(t,f_{0},\dot{f}_{0})=f_{0}t+\dot{f}_{0}t^{2}$, 
where $f_{0}$ and $\dot{f}_{0}$ are the intrinsic frequency and spindown of the source respectively, and $t$ is time. In the case of an actual search for 
GW signals from pulsars, $\Phi_{\alpha j b}(\vec{\lambda})$ will not be so simple. However, this model is sufficiently complex to effectively demonstrate 
all of the approximations to be discussed here. 

In discrete form, $\Phi(t,f_{0},\dot{f}_{0})$ can be written as $\Phi_{\alpha j \beta l}(\gamma)=(\beta+Ml)(N\alpha+j)/NM+\gamma(N\alpha+j)^{2}/N^{2}M^{2}$, 
where the long time baseline frequency index $b=\beta+Ml$. The chosen discretization of the spindown parameter $\dot{f}_{0}\equiv \gamma/T^{2}$ is not 
practically appropriate. However, in an actual search, a grid of points in spindown parameter space will be chosen to ensure an acceptable loss of power 
from unresolved signals. Thus, in the following discussion, only searches for resolved $\dot{f}_{0}$ parameters will be considered.

\begin{description}
\item Approximation 1: By Taylor expanding the model phase function $\Phi(t)$ about the middle of 
each short duration time series (i.e. about $j=N/2$) and discarding terms of order $(j/N)^{2}\equiv t^{2}$ and higher, in the limit 
$N\rightarrow \infty$ the function $P_{\alpha k}(\beta,l,\gamma)$ is Re $P_{\alpha k}(\beta,l,\gamma)=\mbox{sinc }x$ and 
Im $P_{\alpha k}(\beta,l,\gamma)=(1- \cos x)/x$. In the phase model considered here $x=-2\pi \left(\beta/M+l+(2\alpha+1)\gamma/M^{2}-k\right)$ and 
$Q_{\alpha}(\beta,l,\gamma)=\exp\left\{ -2\pi i(\alpha \beta/M +\alpha \gamma^{2}/M^{2})\right\}$. 

\item Approximation 2: Consider the case where the short time baseline is chosen such that the instantaneous model frequency 
$f(t)=\dot{\Phi}(t,f_{0},\dot{f}_{0},\ddot{f}_{0},\ldots)$ does not move by more than one short time baseline frequency bin over the 
duration of a short time baseline data set, i.e. in the model discussed here  $|\dot{f}_{0}|T/M<M/T$. Then for a given $\alpha$, the 
function $P_{\alpha k}(b,\vec{\lambda})$ will be peaked in power about the model frequency averaged over the duration of time associated with 
the $\alpha$th short data set, i.e. about $x=0$ (the first three terms in the above definition of $x$ are the index of this average model 
frequency). Thus only a few terms around this model frequency will contribute significantly to the summation over $k$ in equation \ref{deft}.

\item Approximation 3: The semi--periodic nature of $P_{\alpha k}(b,\vec{\lambda})$ means that this function can be efficiently evaluated from a look--up table 
of values containing the periodic parts, and three further operations: to calculate one instance of $P_{\alpha k}(b,\vec{\lambda})$ will require only 8 floating 
point operations.

\item Approximation 4: If one approximates the model frequency parameter $\beta$ in the calculation of $P_{\alpha k}(\beta,l,\gamma)$ as a fixed value, for 
example with $\beta=\beta_{0}$, equation \ref{deft} can be calculated as an FFT, i.e.
\begin{equation}
\hat{x}_{\beta l}(\gamma)= \sum_{\alpha=0}^{M-1}\left[Q^{'}_{\alpha}(\gamma)\sum_{k}^{n_{term}}\tilde{x}_{\alpha k}P_{\alpha k}(\beta_{0},l,\gamma)\right]
e^{-2\pi i\frac{\alpha\beta}{M}}, \label{deft2}
\end{equation}
where $n_{term}$ relates to approximation 2, $P_{\alpha k}(\beta,l,\gamma)$ is defined above, and for the phase model discussed here 
$Q^{'}_{\alpha}(\gamma) = \exp\left\{-2\pi i\left(\alpha^{2}\gamma/M^{2}-\gamma/4M^{2}\right)\right\}$. Thus for values of $\beta$ 
sufficiently near to $\beta_{0}$, the loss in power due to this approximation will be small. To obtain $\hat{x}_{\beta l}(\gamma)$ for 
other values of $\beta$, the calculation must be repeated using another $\beta_{0}$.
\end{description}

\section*{Results and Discussion}

Numerical tests have shown that if one chooses $10\%$ as an acceptable loss in power in comparison to the optimal case, 
then $N_{FFT}=8$ and $n_{term}=16$ are the preferred parameter combination, if the short time baseline $T/M$ is chosen such that in the phase model 
discussed here $|\dot{f}_{0}|T/M<M/T$. If one decides that only a $5\%$ loss in optimal power is acceptable, 
then this can be achieved with the same parameters, but choosing $T/M$ such that $|\dot{f}_{0}|T/M<M/2T$.

The computational cost of calculating one DeFT in stage I in floating point operations is
\begin{equation}
C_{DeFT} \simeq 5.3\times 10^{10}\left(\frac{B}{300 \mbox{ Hz}}\right)\left(\frac{T}{1 \mbox{ day}}\right)
\left(\frac{N_{FFT}}{8}\right)\left(\frac{n_{term.}}{16}\right),
\end{equation}
where $B$ is the bandwidth of the search. This is comparable to the computational cost of the corresponding steps in the Hierarchical Stack / Slide 
algorithm of Brady and Creighton (``Searching for Periodic Sources with LIGO: Hierarchical Searches'' gr--qc9812014). The Hough--Hierarchical search 
algorithm also has a number of computational advantages. To calculate a given bandwidth of a DeFT requires only the FFT data 
from this bandwidth and an additional small overlap. Thus the algorithm can be easily parallelized by distributing data and work by bandwidth; and no 
communication between processors is required. Also, the complete three stage algorithm can be arranged in such a way that once a bandwidth of FFT data 
is read from disk by a processor, all computation required on this data can be carried out while this data is held in memory, thus time spent reading 
data from disk is a negligible fraction of the total computational time: each processor will need to read roughly 40 Mb from disk once every two weeks. 
Furthermore, little additional memory is required as workspace for stages I and III: less than 100 kb. 

The GEO600 data analysis team are currently working on coding this algorithm in a computationally optimal manner, as well as integrating this with 
the Hough Transform part of the procedure.
\end{document}